\documentclass[prl,superscriptaddress,showpacs,twocolumn]{revtex4-1}
\usepackage{amssymb}
\usepackage{graphicx}
\usepackage{dcolumn}
\usepackage{bm}
\usepackage{amsmath}
\usepackage{epstopdf}

\begin{document}

\title{Chiral Majorana Edge Modes and Vortex Majorana
Zero Modes in Superconducting Antiferromagnetic Topological
Insulator}
\author{Beibing Huang}
\email{hbb4236@ycit.edu.cn}
\affiliation{Department of Physics, Yancheng Institute of Technology, Yancheng, 224051, China}
\author{Xiaosen Yang}
\affiliation{Department of physics, Jiangsu University, Zhenjiang, 212013, China}
\author{Qinfang Zhang}
\affiliation{School of Materials Science and Engineering, Yancheng Institute of Technology, Yancheng 224051, China}
\author{Ning Xu}
\email{nxu@ycit.cn}
\affiliation{Department of Physics, Yancheng Institute of Technology, Yancheng, 224051, China}

\date{\today}

\begin{abstract}
The antiferromagnetic topological insulator (AFTI) is topologically
protected by the combined time-reversal and translational symmetry
$\mathcal{T}_c$. In this paper we investigate the effects of the
$s$-wave superconducting pairings on the multilayers of AFTI, which
breaks $\mathcal{T}_c$ symmetry and can realize quantum anomalous
Hall insulator with unit Chern number. For the weakly coupled
pairings, the system corresponds to the topological superconductor
(TSC) with the Chern number $C=\pm 2$. We answer the following
questions whether the local Chern numbers and chiral Majorana edge
modes of such a TSC distribute around the surface layers. By the
numerical calculations based on a theoretic model of AFTI, we find
that when the local Chern numbers are always dominated by the
surface layers, the wavefunctions of chiral Majorana edge modes must
not localize on the surface layers and show a smooth crossover from
spatially occupying all layers to only distributing near the surface
layers, similar to the hinge states in a three dimensional
second-order topological phases. The latter phase can be
distinguished from the former phase by the measurements of the local
density of state. In addition we also study the superconducting vortex phase transition in this system and find that the exchange field in the AFTI not only enlarges the phase space
of topological vortex phase but also enhances its topological stability.  These conclusions will stimulate the investigations
on superconducting effects of AFTI and drive the studies on chiral
Majorana edge modes and vortex Majorana zero modes into a new era.
\end{abstract}

\maketitle

\section{Introduction}

The highly effectiveness of quantum computation is rooted in its
parallel properties ensured by the superposition principles of quantum
mechanism. However the fragileness of quantum systems easily causes
random errors in the computational process, which can be overcome by
encoding all information into nonlocal qubits (topological quantum
computation, TQC) \cite{kitaevtqc, nayak}. Majorana fermions
\cite{majorana}, realized as the Majorana zero modes in the
superconducting vortex \cite{readgreen, fukane, hosur, jia1, jia2,
xugang, ding1, ding2, feng1, feng2} or the chiral Majorana edge
modes \cite{qi, wangjin, jau, alicea} in the topological
superconductors (TSC), can be used to construct topological qubits.
However, the computational processes of TQC require braiding the
Majorana fermions \cite{nayak, ivan}, which is a very difficult task
until now for the Majorana zero modes in the superconducting vortex
cores. By contrast, the chiral Majorana edge modes can interchange
themselves in the transport process and are more easily manipulated
\cite{lianbiao}.

In 2010, Qi etc suggested that chiral Majorana edge modes can be
realized as the edge states in the heterostructures of s-wave
superconduting proximity to a quantum anomalous Hall insulator
(QAHI) \cite{qi}. But the experiments in this direction \cite{e1,
e2, e3} arouse much debates about the complicated magnetic disorder
effects in the QAHI obtained by magnetically doping the topological
insulator \cite{disorder1, disorder2}. The appearance of the
antiferromagnetic topological insulator (AFTI) finds a way out of
this dilemma to a large extent \cite{moore}. In the AFTI, taking the
characteristic Van der Waals layer material $\text{MnBi}_2\text{Te}_4$ for example,
the magnetic exchanges show intralayer (interlayer) ferromagnetic
(antiferromagnetic) structures with the moments perpendicular to the
layer plane, and the topological properties are protected by the
combined time-reversal and lattice translational symmetry
\cite{mnbite0, mnbite, mnbite1, mnbite2, mnbite3}. For an odd (even)
number of multilayers, the average moment is nonzero (zero) and the
system is an QAHI (axion insulator) with unit (zero) Chern number.
The related experiments have observed quantized Hall plateau without
the external magnetic field at the high temperature \cite{qah1,
qah2, qah3}. These characters such as stoichiometric purity and Van
der Waals structure hint that the slab of AFTI can be used as one of
potential candidates to realize the chiral Majorana edge modes.

The multilayers of AFTI have two surfaces and show the gapped
surface states. These surface states originate from gapping out the
gapless ones of topological insulator and correspond to a half QAHI
\cite{qi2, fu2}, described by a massive Dirac fermion. Moreover the
calculation of local Chern number shows that the contribution to
Chern number comes from surface layers, and as one goes deeper, the
local Chern number oscillates around zero \cite{lcn1, lcn2, lcn3}.
According to Qi's suggestion \cite{qi}, once the weakly coupled
$s$-wave pairings are introduced, the multilayers of AFTI with odd
number layers become the TSC with Chern number $C=2$. In this paper,
we ask the following two questions. The first one is whether the
local Chern numbers of such TSC with $C=2$ distribute around the
surface layers. If the answer is positive, the next question is
whether the chiral Majorana edge modes also distribute around the
surface layers. By the numerical calculations based on a theoretic
model of AFTI, we find that even if the surface Chern numbers
dominate, the wavefunctions of chiral Majorana edge modes must not
localize near the surface layers and show a smooth crossover from
spatially occupying all layers to only occupying surface layers when
some parameters are adjusted. In addition we also study the superconducting vortex phase transition in this system and find that the exchange field in the AFTI not only enlarges the phase space
of topological vortex phase but also enhances its topological stability. These conclusions will stimulate the
investigations on the superconducting effects of AFTI.

In this paper, we are interested in the TSC with Chern number $C=2$
realizable in superconducting multilayers of AFTI, which are out of
spotlight in contrast to the TSC with $C=1$. The plan of this paper
is as follows. In the section 2, starting from the model of
topological insulator, we give the tight-binding model of
superconducting multilayer of AFTI by incorporating
the antiferromagnetic exchange field in $z$ direction and $s$-wave
on-site intraorbital pairings. In the section 3 we map out the phase
diagrams of the system and analyze the distribution of Chern number
to exemplify its local properties. We also find that the chemical
potential plays an important role in determining available
topological phases in our suggested model. The section 4 is devoted
to study the wavefunctions of chiral Majorana edge modes. By
tracking the evolutions of wavefunctions, we find that the edge
modes can continuously evolve into the state only occupying surface
layers. We illustrate this crossover from the structure of bulk
energy bands. In the section 5, the effects of the antiferromagnetic exchange field on the superconducting vortex phase transition are studied. Finally a brief conclusion is made in the section 6.

\section{Model}

We start with a four-band model for topological insulators on a
cubic lattice with two orbitals and spins per cubic site \cite{ti}
\begin{eqnarray}
H_{TI}(k_x,k_y,k_z)&=&[m+t(\cos{k_x}+\cos{k_y}+\cos{k_z})]\tau_z\nonumber\\&+&v
(\sin{k_x}\sigma_x+\sin{k_y}\sigma_y+\sin{k_z}\sigma_z)\tau_x,\nonumber\\
\label{1}
\end{eqnarray}
where $\sigma_{x,y,z}$ and $\tau_{x,y,z}$ are two sets of Pauli
matrices acting on spin and orbital spaces respectively. This
Hamiltonian is in the strong topological insulator phase protected
by the time-reversal symmetry with
$\mathcal{T}=-i\sigma_y\mathcal{K}$ if $|t|<|m|<3|t|$ and $v \ne 0$,
where $\mathcal{K}$ is the complex conjugation operator.

The Hamiltonian of AFTI can be achieved by adding the
antiferromagnetic exchange term along $z$ direction into the
Hamiltonian (\ref{1}) \cite{moore, xuyong}. This exchange term
breaks the translation symmetry in $z$ direction and magnifies the
unit cell of the topological insulator to include four orbitals.
Choosing the magnetic moment colinear with $z$ direction, AFTI is
described by the Hamiltonian
\begin{eqnarray}
H_{AFTI}(k_x,k_y,k_z)&=&[m+t(\cos{k_x}+\cos{k_y})]\tau_z+\Gamma\varrho_z\sigma_z\nonumber\\
&+&\varrho_x[t\cos{(k_z/2)}\tau_z+v\sin{(k_z/2)}\sigma_z\tau_x]\nonumber\\&+&v
(\sin{k_x}\sigma_x+\sin{k_y}\sigma_y)\tau_x, \label{2}
\end{eqnarray}
where the Pauli matrices $\varrho_{x,y,z}$ act on the bilayer space
induced by antiferromagnetic structure and $\Gamma$ is the
antiferromagnetic exchange field. The model (\ref{2}) has the
combined time-reversal and translational symmetry
$\mathcal{T}_c=e^{-ik_z/2}\rho_x\mathcal{T}$ and its topological
properties are decided by $Z_2$ number at $k_z=0$ plane. We find
that as long as the gap from topological insulator is not closed
when $\Gamma$ is increased, the model (\ref{2}) is an AFTI. In
Fig.1(a) and (b), we show the surface states of the model (\ref{2}),
consistent with the our expectations.

\begin{figure}[tbp]
\centering
\includegraphics[width=0.45\textwidth]{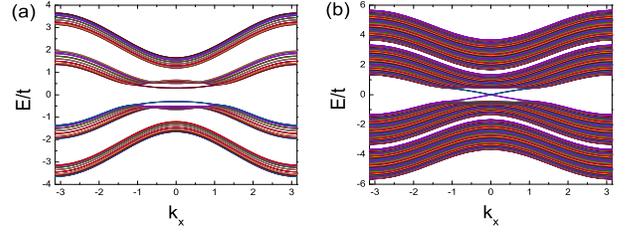}
\caption{The excitation spectra in the surface Brillouin zones
$k_xk_y$ with $k_y=0$ (a) and $k_xk_z$ with $k_z=0$ (b) for the AFTI
described by the model (\ref{2}). In (a) [(b)] the surfaces are
ferromagnetic (antiferromagnetic) and breaks (conserves) the
combined time-reversal and translational symmetry $\mathcal{T}_c$,
thus the surface states correspond to a gapped (gapless) Dirac
fermion. The parameters are $m/t=-2.5$, $v/t=0.5$, $\Gamma/t=0.6$.}
\label{fig.1}
\end{figure}

In this paper, we consider a multilayer from the AFTI, which is
finite length $L_z$ along $z$ direction. At the same time, we also
assume that the multilayer is superconducting with $s$-wave on-site
intraorbital pairings. Without loss of generality, we assume all
pairings are equal, denoted by $\Delta$. Further introducing Pauli
matrices $s_{x,y,z}$ acting on the particle-hole space, our
Hamiltonian is
\begin{eqnarray}
H_{SML}(k_x,k_y)&=&[m+t(\cos{k_x}+\cos{k_y})]s_z\tau_z-\mu s_z\nonumber\\&+&v\sin{k_x}\sigma_x\tau_x+v
\sin{k_y}s_z\sigma_y\tau_x+\Delta
s_y\sigma_y\nonumber\\&+&\frac{t}{2}s_z\widetilde{\varrho}_x\tau_z+\frac{v}{2}\widetilde{\varrho}_y\sigma_z\tau_x+\Gamma
s_z\widetilde{\varrho}_z\sigma_z, \label{3}
\end{eqnarray}
where we have added the chemical potential $\mu$. The new $L_z
\times L_z$ matrices $\widetilde{\varrho}_{x,y,z}$ are used to
describe the interlayer coupling, which are defined as follows
\begin{eqnarray}
\widetilde{\varrho}_{x,y}=\left( \begin{array}{cccccc} 0& a & 0&...\\
a^*& 0 & a&...\\ 0& a^* & 0&...\\ ...& ...&
...&...\end{array}\right),
\widetilde{\varrho}_{z}=\left( \begin{array}{cccccc} 1& 0 & 0&...\\
0& -1 & 0&...\\ 0& 0 & 1&...\\ ...& ...& ...&...\end{array}\right)
\end{eqnarray}
with $a=1$ for $\widetilde{\varrho}_{x}$ and $a=-i$ for
$\widetilde{\varrho}_{y}$.

The model (\ref{3}) is our starting point of this paper. We choose
the model parameters to ensure that the gap closing of the system
happens at $k=0$. We set
$m/t=-2.5$, $v/t=0.5$.

\begin{figure}[tbp]
\centering
\includegraphics[width=0.45\textwidth]{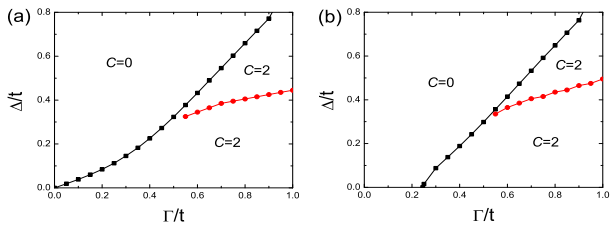}
\caption{The phase diagrams for the multilayer model (\ref{3}) as
the function of the antiferromagnetic exchange field $\Gamma$ and
superconducting gap $\Delta$ with the chemical potential $\mu/t=0$
(a) and $\mu/t=0.1$ (b). The lines with squares separate the normal
superconducting phase from the TSC with Chern number $C=2$; while
the lines with circles differentiate two phases in the TSC. These
two phases are defined depending on whether the chiral Majorana edge
modes only occupy a few layers near the surfaces of the multilayer.
The triangle-like regions correspond to such localized phase. It
should be noted that in (b) there exists a small phase space not
shown for TSC with $C=1$ between the phases with $C=2$ and $C=0$.}
\label{fig.2}
\end{figure}

\section{Phase Diagram}

We construct the phase diagram by tracking the gap closings and the
Chern number of the system \cite{fukui}
\begin{eqnarray}
C=\frac{1}{2\pi}\int dk_xdk_y \text{Tr} \hat{\Omega}_{k,xy} \label{7}
\end{eqnarray}
where $\hat{\Omega}_{k,xy}$ are the Berry curvature matrix with
\begin{eqnarray}
\hat{\Omega}_{k,xy}^{nn'}=\nabla_k \times
<\psi_{nk}|i\nabla_k|\psi_{n'k}>
\end{eqnarray}
and $|\psi_{nk}>$ is the eigenstates of the occupied bands with
eigenvalues $E_{nk}$ in ascending order. We study the Chern number
as the function of the antiferromagnetic exchange field $\Gamma$ and
superconducting gap $\Delta$. The phase diagrams for two different
chemical potentials are shown in Fig.2.

For $\mu=0$, the Hamiltonian (\ref{3}) has an emergent symmetry $\Xi
H_{SML}(k_x,k_y) \Xi^{-1}=H_{SML}(k_x,k_y)$ with
$\Xi=s_y\sigma_y\tau_y$. This symmetry transforms the particles in
an orbital into holes in the other orbital on the same layer but
keeps the momentum and spin invariant. Therefore, if $\psi_{nk}$ is
an eigenstate of $H_{SML}(k_x,k_y)$, $\Xi \psi_{nk}$ is also the
eigenstate with the same energy. We have numerically checked that
these two eigenstates are not linearly correlated. Thus for $\mu=0$,
all energy bands are double degenerate. This result directly
influences the available phases in the model (\ref{3}). Generally
the topological phase transition accompanies the gap closings and
the dispersion around the phase transition point are gapless Dirac
type. Before and after the phase transition, the Dirac mass is
inverted and for every this transition, the Chern number of the
system changes $\pm 1$ \cite{bissard}. Now for $\mu=0$, the double
degeneracy means that the Chern number changes in unit of $\pm 2$.
Since a weakly superconducting coupled QAHI with $C=1$ realizes a
TSC with $C=2$, thus our system can only show the even Chern number
states at $\mu=0$.

When $\mu\neq 0$, the symmetry $\Xi$ and double degeneracy are
broken, thus the Chern number changes in unit of $\pm 1$ and we can
observe the TSC with $C=1$. But in terms of parameters we have
chosen, this phase space is very small, so that we do not show this
phase region in the phase diagram Fig.2(b). In Fig.2(b), when
$\Gamma$ is small, the system is a normal superconductor. In this
region, the normal state is a metal, not a QAHI. With the increase
of the exchange field $\Gamma$, the gap of the normal state
increases. When the gap of the normal state is beyond the chemical
potential, the system enters into the topological phase. A finite
chemical potential reduces the regions of TSC. This conclusion is
consistent with the conventional wisdom.

The Chern number used to differentiate the different phases is
defined for the whole multilayer. To study the distribution of Chern
number as the function of layer index of multilayer, we need find
how different layers contribute to $C$. This local Chern number
$C_l$ for the $l$th layer can be calculated from the formula
\cite{lcn1, lcn2, lcn3}
\begin{eqnarray}
C_l=\frac{1}{2\pi}\int dk_xdk_y \text{Tr}
[\hat{\Omega}_{k,xy}\hat{P}_{k,l}] \label{9}
\end{eqnarray}
by inserting a projection $\hat{P}_{k,l}$ onto layer $l$
\begin{eqnarray}
\hat{P}_{k,l}^{nn'}=\sum_{j\in l} \psi_{nk,j}^*\psi_{n'k,j},\label{10}
\end{eqnarray}
where the summation $j\in l$ is done if the $j$th component of the
eigenfunction $\psi_{nk}$ represents the degree of freedom in layer
$l$. Since $\sum_l\hat{P}_{k,l}^{nn'}=\delta_{n,n'}$, we have
$\sum_l C_l=C$. The formalism of $C_l$ was developed for calculating
the full surface anomalous Hall conductivity in a multilayer of
magnetoelectric insulator.

\begin{figure}[tbp]
\centering
\includegraphics[width=0.45\textwidth]{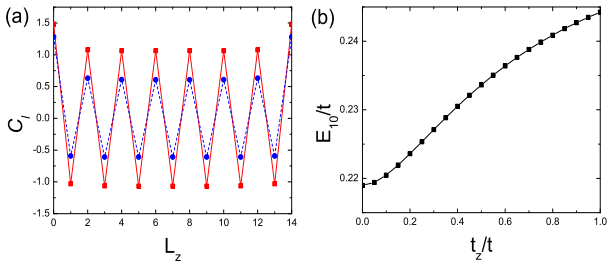}
\caption{(a) The local Chern number $C_l$. With the increase of the
superconducting gap, the oscillation of the local Chern number shows
decayed amplitude. (b) The evolution of first excitation level
$E_{1k}$ at $k=0$ when the hopping in $z$ direction decreases to
zero. The spin-orbit coupling in $z$ direction plays no roles for
the gap of the system at the time-reversal invariant momenta. In (a)
the parameters for the red solid (blue dashed) line with squares
(circles) are $\Gamma/t=1.0$ and $\Delta/t=0.02$ ($\Delta/t=0.65$);
In (b) $\Gamma/t=1.0$, $\Delta/t=0.6$.} \label{fig.3}
\end{figure}

Below we directly apply this quantity to our case. Fig.3 shows $C_l$
for some different parameters using a $10\times 10$ sampling in the
Brillouin zone. In all cases, we find that $C_l$ on the two surface
layers approach half of the Chern number of the superconducting
multilayer, and as one goes deeper into the interior, $C_l$
oscillates around zero with the average of neighbor layers
approaching zero. Therefore the contribution to the whole Chern
number comes from a few layers near the surfaces. These results for
$C_l$ can be qualitatively understood as follows. We find that the
gap of the topological phase in Fig.2 do not close when the hopping
in $z$ is adjusted to zero (Fig.3(b)); In addition the spin-orbit
coupling in $z$ direction does not influence the gap of the system
at the time-reversal invariant momenta. Therefore from the
perspective of topology, it is very appropriate to consider the
model (\ref{3}) adiabatically connected to the model without
interlayer couplings. In this limit, every independent layer can be
regarded as the TSC with $C=\pm 1$ depending on the sign of the
antiferromagnetic exchange field $\Gamma$. As a result the Chern
numbers of the interior layers average to zero, while surface layers
at either end of the multilayer are not completely canceled,
realizing a TSC with $C=\pm 1$ at both surfaces.

\section{Majorana Wavefunctions}

In this paper, we are interested in the TSC with $C=2$. In Fig.3, we
have seen that the local Chern number $C_l$ mainly distribute on the
surface layers with the averages of the interior layers approaching
zero. Now we want to consult that when the Chern number concentrates
on the surface layers, whether the chiral Majorana edge modes also
concentrates on the surface layers. Below we numerically calculate
the wavefunctions of chiral Majorana edge modes to illustrate this
question, by assuming an open boundary condition in $y$ direction. Fig.4 shows some characteristic cases for Majorana
edge modes.

\begin{figure}[tbp]
\centering
\includegraphics[width=0.45\textwidth]{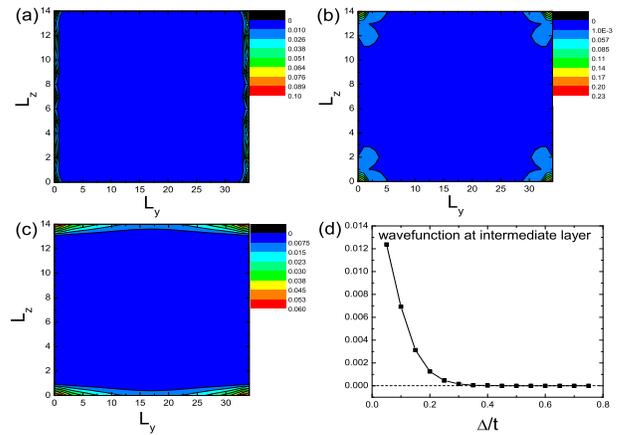}
\caption{The evolutions of wavefunctions for the chiral Majorana
edge modes as the superconducting gap $\Delta$ is adjusted. The
parameters are $\Gamma/t=1.0$ and $\Delta/t=0.02$, $0.65$, $0.80$ in
(a)-(c).} \label{fig.4}
\end{figure}

The normal state of our model is a QAHI and always gapped if we
decrease the exchange fields $\Gamma$ on the interior layers to zero
but keep them invariant on the two surface layers, thus its chiral
edge modes are trapped on the whole side surfaces \cite{qi2}. On the
other hand the TSC with $C=2$ in the Fig.2 can be continuously
connected to the normal state, we naturally expect that when the
superconducting gap $\Delta$ is small, the chiral Majorana edge
modes will occupy all layers of the multilayer, which is consistent
with the results in Fig.4(a). While for a big superconducting gap
$\Delta$, the wavefunctions are absent from the interior layers and
localized near surface layers (Fig.4(b)). This distribution of
wavefunctions is similar to the hinge states in a three dimensional
second-order topological phases \cite{hinge1, hinge2, hinge3,
hinge4}. Between these two limits, there is a
smooth crossover of wavefunction evolution. From this perspective,
we can differentiate two kinds of phases of TSC with $C=2$. In order
to discriminate these two phases, we defined the wavefunctions at
intermediate layer of chiral Majorana edge modes equaling to zero as
the phase transition point between these two phases. Such behaviors
as the function of superconducting gap $\Delta$ are shown in
Fig.4(d). We have found some phase transition points shown in Fig.2
as the circles for different parameters. For small $\Gamma$, the
search for the critical point is costly, so we only make some
calculations for large $\Gamma$. Here we note that such two phases
are topologically equivalent with each other, in spite of the
differences on the wavefunctions.

\begin{figure}[tbp]
\centering
\includegraphics[width=0.45\textwidth]{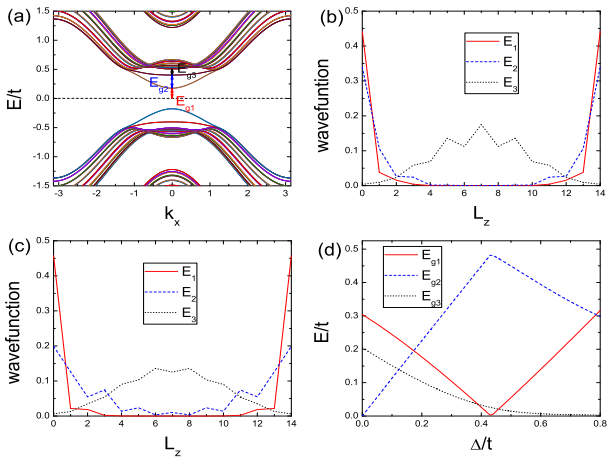}
\caption{(a) The excitation spectrum of the multilayer for $k_y=0$.
Three excitation gaps $E_{g1}=E_{10}$, $E_{g2}=E_{20}-E_{10}$,
$E_{g3}=E_{30}-E_{20}$ at $k=0$ are schematically illustrated;
(b)-(c) The wavefunctions of three lowest energy levels $E_{ik}$ at
$k=0$ as the function of layer index; (d) The evolutions of three
excitation gaps as the function of superconducting pairings. The
parameters are $\Delta/t=0.2$ in (a), $\Delta/t=0.1$ in (b),
$\Delta/t=0.4$ in (c), $\Gamma/t=0.6$.} \label{fig.5}
\end{figure}

The TSC with $C=2$ will transit into another phase when the
superconducting gap is added further. For $\mu=0$, the system
transits into the trivial phase. In this process of phase
transition, the chiral Majorana edge modes have the same
localization length along the $y$ direction. Increasing the
superconducting gap, the localization length gradually increases
(Fig.4(c)) and even becomes divergent at the critical point, so that
edge modes merge into the bulk states in $y$ direction, leaving a
trivial gapped superconductor on the other side of the transition.
Similarly for $\mu\neq 0$, there is phase transition to TSC with
$C=1$. The nonzero chemical potential leads to independent evolution
of edge modes. At the critical point, one edge mode becomes bulk
state while the other is still invariant. But it is worthwhile to
note that, the chiral Majorana edge modes always localized near the
surface layers in these phase transitions. This property is always
invariant.

From the distributions of wavefunctions of chiral Majorana edge
modes, we find that although the Chern number is mainly decided by
the surface layers, generally the wavefunctions of edge modes do not
only occupy the surface layers. In order to illustrate localization
to the surface layers, we observe the evolution of bulk excitation
spectrum $E_{nk}$ at $k=0$, since as stated before we have chosen
the model parameters to ensure that the gap closing of the system
happens at this point. The excitation spectrum of the multilayer
consists of two parts. The first (second) part comes from gapless
surface states (gapped bulk states) of topological insulator. When
imposed on the staggered magnetic moments and small superconducting
pairings, these states still occupied surface layers (bulk)
(Fig.5(b)), contributing a large local Chern number to the whole
Chern number. Further increasing the superconducting pairings,
although the gaps of the system $E_{g1}=E_{10}$ and
$E_{g3}=E_{30}-E_{20}$ decrease, the gap $E_{g2}=E_{20}-E_{10}$
gradually increases (Fig.5(d)), not only leading to the mixture of
surface energy level $E_{2k}$ into the bulk ones (Fig.5(c)) but also
that the low energy properties of the system is only decided to a
larger extent by the first excitation level $E_{1k}$ belonging to
surface layers. Thus once a boundary is imposed, we expect the
chiral Majorana edges modes will also localize near the surface
layers. This illustrates our observations on the wavefunctions.

\begin{figure}[tbp]
\centering
\includegraphics[width=0.45\textwidth]{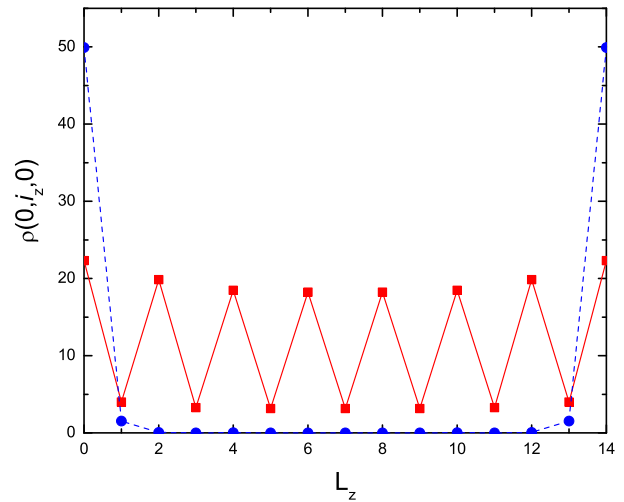}
\caption{The local zero-energy density of state $\rho(i_y=0, i_z,
\omega)$ at $\omega=0$. In order for calculation, we have added a
small imaginary part $\delta=10^{-3}$ to $\omega$. The local
zero-energy density of state $\rho(i_y=L_y-1, i_z, 0)$ shows the
same structure. The parameters for the red solid (blue dashed) line
with squares (circles) are $\Gamma/t=1.0$ and $\Delta/t=0.02$
($\Delta/t=0.65$).} \label{fig.6}
\end{figure}

Two kinds of TSC with $C=2$ have the common bulk properties, however
the occupation of the chiral Majorana edge modes near the surface
layers should leave the fingerprints on the local low-energy density
of state, which can be calculated from the retarded Green function
$\hat{G}_{ret}(k_x,\omega)$ for the system with edges $\rho(i_y,
i_z, \omega)=-\sum_{ik_x}\text{Im} [\hat{G}_{ret}^{ii}(k_x,\omega)]$, where
the summation for $i$ concerns the all degree of freedom on the
lattice site $(i_y,i_z)$ and the label $\text{Im}[\cdot\cdot\cdot]$ means
to take the imaginary part. Fig.6 presents the featured local
zero-energy density of state $\rho(i_y=0, i_z, 0)$ on the side,
which has the similar behaviors with the wavefuntions of the chiral
Majorana edge modes. Physically $\rho(i_y=0, i_z, 0)$ can be
measured as the zero-bias conductance peak using the scanning
tunneling spectroscopy and can be used as the smoking gun to
discriminate these two phases experimentally.

\section{Vortex Phase Transition}

For the topological insulator, a superconducting vortex line will
bind two Majorana zero modes, which are localized near the
intersections of two surfaces and vortex line when the chemical
potential $\mu$ is around the charge-neutral point; while tuning
$\mu$ into the bulk states, a quantum vortex phase transition occurs
and two Majorana zero modes disappear \cite{hosur}. The similar
vortex phase transitions have been investigated in superconducting
Dirac and Weyl semimetal \cite{yan}. In our model, the
antiferromagnetic exchange field $\Gamma$ gap out the surface Dirac
point and provides a new knob to manipulate the vortex. Below we
will explore the effects of $\Gamma$ on vortex phase transition.

\begin{figure}[tbp]
\centering
\includegraphics[width=0.45\textwidth]{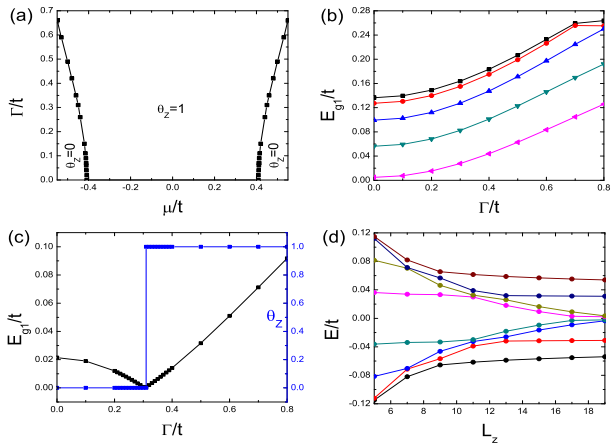}
\caption{(a) The phase diagram for one-dimensional vortex line in
the superconducting AFTI. The region with Zak phase $\theta_Z=1$
($\theta_Z=0$) is nontrivially (trivially) topological. For
$\Gamma=0$, two critical points of the vortex phase transition are
located at $\mu_c/t=\pm 0.41$. (b) The excitation gap $E_{g1}$ as
the function of $\Gamma$ for the different chemical potential. From
top to bottom $\mu/t=0.0$, $0.1$, $0.2$, $0.3$, $0.4$ respectively.
(c) is the same as (b) except $\mu/t=0.45$. In this case the
excitation shows gap closing and reopening, signifying a vortex
phase transition. The change of Zak phase $\theta_Z$ is also
presented. (d) The scaling behavior of some low-energy eigenvalues
for a cubic sample with open boundary condition in all three
directions for $\mu/t=0.1$ and $\Gamma/t=0.7$. We only show the
results for odd number multilayers. Note that for $\mu=0$, the
emergent symmetry $\Theta$ leads to the double of Majorana zero
modes. In all figures $\Delta_0/t=0.4$, $\xi=0.5$.} \label{fig.7}
\end{figure}

In order to model a superconducting vortex line in $z$ direction, we
assume the superconducting gap $\Delta$ in the model (\ref{3})
acquires the spatial dependence
$\Delta_{i_x,i_y}=\Delta_0\tanh{(\sqrt{i_x^2+i_y^2}/\xi)}e^{i\theta}$,
where the phase $\theta=\arctan{(i_y/i_x)}$ reflects the vortex
structure and $\xi$ is the superconducting coherence length. We also
assume the extreme type II limit to neglect the effect of the
magnetic field used to generate the vortices. The vortex line
conserves the translational symmetry in the $z$ direction, so that
$k_z$ is still a good quantum number and our system can be viewed as
a one-dimensional superconductor. However, the complex
superconducting gap breaks the combined time-reversal and
translational symmetry $\mathcal{T}_c$ and our system is
characterized by a well-defined Zak phase $\theta_z$ if the energy
spectrum is fully gapped \cite{xugang, jpsj, budich}. The
topological phase diagram from this calculation is shown in
Fig.7(a). For $\Gamma=0$, the system has a nontrivial topological
region around the charge-neutral point $\mu=0$, consistent with
others results \cite{hosur}. With the increase of
 $\Gamma$, the topological region is gradually enlarged. On the other
hand, we also find the excitation gap nontrivially increases as the
function of $\Gamma$ for the fixed chemical potential in the
topological region (Fig.7(b)), further leading to the robustness of
the ground state. These two points signify that the extra
controllable knob of the antiferromagnetic exchange field $\Gamma$
is very advantageous to enhance the topological stability. For the
chemical potential beyond the critical points at $\Gamma=0$ the
system shows gap closing, where the sudden jump of Zak phase
$\theta_Z$ happens (Fig.7(c)). To further support the results above,
we also diagonalize the Hamiltonian for a cubic geometry with open
boundary condition in all three directions. In Fig. 7(d), we present
some eigenenergies near zero energy for $\mu/t=0.1$ and
$\Gamma/t=0.7$, their scalings with the system size clearly
demonstrate the existence of vortex Majorana zero modes.

For the chemical potential $\mu=0$, as stated before, there exists
an emergent symmetry $\Xi$. Thus the Hamiltonian can be partitioned
into two subspaces related to this symmetry. Considering that the
commutation between this symmetry and the particle-hole symmetry,
every subspace has also the particle-hole symmetry, thus in every
subspace we can also define Zak phase. These two Zak phases is
nontrivial for $\mu=0$. In other words, the emergent symmetry
$\Theta$ leads to the double of Majorana zero modes at $\mu=0$. This
can be used to illustrate near degeneracy between Majorana zero
modes and low energy excitation in Fig.7(d).

\section{Conclusions}

In this paper all calculations have been done for the odd number of
multilayers. However the conclusions about the chiral Majorana
wavefunctions and vortex phase transition are also applicable to the
even number of multilayers. In conclusion we have investigated the
effects of the $s$-wave superconducting pairings on the multilayers
of AFTI, which breaks $\mathcal{T}_c$ symmetry and can realize QAHI
with unit Chern number. For the weakly coupled pairings, the system
corresponds to the TSC with the Chern number $C=\pm 2$. By the
numerical calculations based on a theoretic model of AFTI, we
answered the following questions whether the local Chern numbers and
chiral Majorana edge modes of such a TSC distribute around the
surface layers. On one hand we find that the local Chern number
$C_l$ on the two surface layers approach half of the Chern number of
the superconducting multilayer, and as one goes deeper into the
bulk, $C_l$ oscillates around zero with the average of neighbor
layers approaching zero. Therefore the contribution to the whole
Chern number comes from a few layers near the surfaces; On the other
hand the wavefunctions of the chiral Majorana edge modes must not
localize on the surface layers and show a smooth crossover from
spatially occupying all layers to only distributing near the surface
layers. These two kinds of topological phases can be distinguished
by the measurements of the local density of state. We also discuss
the vortex phase transition in this system and find that the
antiferromagnetic exchange field not only enlarges the phase space
of topological phase but also enhance its the topological stability.
Our conclusion will stimulate the investigation on superconducting
effects of AFTI and drive the studies on chiral Majorana edge modes
into a new era.

\section*{Acknowledgement}

We thank Ming Gong for stimulating discussions. B. H. and X. Y. are supported respectively by National Natural
Science Foundation of China under Grant No. 11547047 and No.
11504143.


\begin{thebibliography}{99}

\bibitem{kitaevtqc} A. Y. Kitaev, Ann. Phys. 303, 2 (2003).
\bibitem{nayak} C. Nayak, S. H. Simon, A. Stern, M. Freedman and S. Das
Sarma, Rev. Mod. Phys. 80, 1083 (2008).
\bibitem{majorana} E. Majorana, Nuovo Cimento 5, 171 (1937).
\bibitem{readgreen} N. Read and D. Green, Phys. Rev. B 61, 10267
(2000).
\bibitem{fukane} L. Fu and C. L. Kane, Phys. Rev. Lett. 100, 096407
(2008).
\bibitem{hosur} P. Hosur, P. Ghaemi, R. S. K. Mong and A. Vishwanath,
Phys. Rev. Lett. 107, 097001 (2011).
\bibitem{jia1} J.-P. Xu, C. Liu, M.-X. Wang, J. Ge, Z.-L. Liu, X. Yang, Y. Chen, Y. Liu, Z.-A. Xu,
C.-L. Gao, D. Qian, F.-C. Zhang and J.-F. Jia, Phys. Rev. Lett. 112,
217001 (2014).
\bibitem{jia2} H.-H. Sun, K.-W. Zhang, L.-H. Hu, C. Li, G.-Y. Wang, H.-Y. Ma, Z.-A. Xu, C.-L. Gao,
D.-D. Guan, Y.-Y. Li, C. Liu, D. Qian, Y. Zhou, L. Fu, S.-C. Li,
F.-C. Zhang and J.-F. Jia, Phys. Rev. Lett. 116, 257003 (2016).
\bibitem{xugang} G. Xu, B. Lian, P. Tang, X.-L. Qi and S.-C. Zhang, Phys. Rev. Lett. 117,
047001 (2016).
\bibitem{ding1} P. Zhang, K. Yaji, T. Hashimoto, Y. Ota, T. Kondo, K. Okazaki, Z. Wang, J. Wen,
G. D. Gu, H. Ding and S. Shin, Science 360, 182 (2018).
\bibitem{ding2} D. Wang, L. Kong, P. Fan, H. Chen, S. Zhu, W. Liu, L. Cao, Y. Sun, S. Du, J.
Schneeloch, R. Zhong, G. Gu, L. Fu, H. Ding and H.-J. Gao, Science
362, 333 (2018).
\bibitem{feng1} Q. Liu, C.Chen, T.Zhang, R. Peng, Y.-J. Yan, C.-H.-P.Wen, X.Lou, Y.-L. Huang,
J.-P. Tian, X.-L. Dong, G.-W. Wang, W.-C. Bao, Q.-H. Wang, Z.-P.
Yin, Z.-X. Zhao and D.-L. Feng, Phys. Rev. X 8, 041056 (2018).
\bibitem{feng2} C. Chen, Q. Liu, T.-Z. Zhang, D. Li, P.-P. Shen, X.-L. Dong, Z.-X. Zhao, T.
Zhang and D.-L. Feng, Chin. Phys. Lett. 36, 057403 (2019).
\bibitem{qi} X.-L. Qi, T. L. Hughes, S.-C. Zhang, Phys. Rev. B 82, 184516 (2010).
\bibitem{wangjin} J. Wang, Q. Zhou, B. Lian and S.-C. Zhang, Phys. Rev. B
92, 064520 (2015).
\bibitem{jau} J. D. Sau, R. M. Lutchyn, S. Tewari and S. Das Sarma,
Phys. Rev. Lett. 104, 040502 (2010).
\bibitem{alicea} J. Alicea, Phys. Rev. B 81, 125318 (2010).
\bibitem{ivan} D. A. Ivanov, Phys. Rev. Lett. 86, 268 (2001).
\bibitem{lianbiao} B. Lian, X.-Q. Sun, A. Vaezi, X.-L. Qi and S.-C. Zhang, Proc.
Natl. Acad. Sci. 115, 10938 (2018).
\bibitem{e1} Q. L. He, L. Pan, A. L. Stern, E. C. Burks, X. Che, G. Yin, J. Wang, B. Lian, Q.
Zhou, E. S. Choi, K. Murata, X. Kou, Z. Chen, T. Nie, Q. Shao, Y.
Fan, S.-C. Zhang, K. Liu, J. Xia and K. L. Wang, Science 357, 294
(2017).
\bibitem{e2} J. Shen, J. Lyu, J. Z. Gao, Y.-M. Xie, C.-Z. Chen, C. Cho, O. Atanov, Z. Chen, K.
Liu, Y. J. Hu, K. Y. Yip, S. K. Goh, Q. L. He, L. Pan, K. L. Wang,
K. T. Law and R. Lortz, Proc. Natl. Acad. Sci. 117, 238 (2020).
\bibitem{e3} M. Kayyalha, D. Xiao, R. Zhang, J. Shin, J. Jiang, F. Wang, Y.-F. Zhao, R. Xiao,
L. Zhang, K. M. Fijalkowski, P. Mandal, M. Winnerlein, C. Gould, Q.
Li, L. Molenkamp, M. H. W. Chan, N. Samarth and C.-Z. Chang, Science
367, 64 (2020).
\bibitem{disorder1} W. Ji and X.-G. Wen, Phys. Rev. Lett. 120, 107002 (2018).
\bibitem{disorder2} Y. Huang, F. Setiawan and J. D. Sau, Phys. Rev. B 97, 100501 (2018).
\bibitem{moore} R. S. K. Mong, A. M. Essin and J. E. Moore, Phys. Rev. B 81, 245209 (2010).
\bibitem{mnbite0} M. M. Otrokov et al., Nature (London) 576, 416 (2019).
\bibitem{mnbite} M. M. Otrokov, I. P. Rusinov, M. Blanco-Rey, M. Hoffmann,
A. Y. Vyazovskaya, S. V. Eremeev, A. Ernst, P. M. Echenique, A.
Arnau and E. V. Chulkov, Phys. Rev. Lett. 122, 107202 (2019).
\bibitem{mnbite1} Y. Gong, J. Guo, J. Li, K. Zhu, M. Liao, X. Liu, Q. Zhang, L. Gu, L. Tang, X. Feng,
D. Zhang, W. Li, C. Song, L. Wang, P. Yu, X. Chen, Y. Wang, H. Yao,
W. Duan, Y. Xu, S.-C. Zhang, X. Ma, Q.-K. Xue and K. He, Chin. Phys.
Lett. 36, 076801 (2019).
\bibitem{mnbite2} D. Zhang, M. Shi, T. Zhu, D. Xing, H. Zhang and J. Wang, Phys. Rev. Lett. 122, 206401 (2019).
\bibitem{mnbite3} J. Li, Y. Li, S. Du, Z. Wang, B.-L. Gu, S.-C. Zhang, K. He, W. Duan and Y. Xu,
Sci. Adv. 5, eaaw5685 (2019).
\bibitem{qah1} Y. Deng, Y. Yu, M. Z. Shi, Z. Guo, Z. Xu, J. Wang, X. H. Chen and Y.Zhang, Science 367, 895 (2020).
\bibitem{qah2} C. Liu, Y. Wang, H. Li, Y. Wu, Y. Li, J. Li, K. He, Y. Xu, J. Zhang and Y. Wang,
Nat. Mater. 19, 522 (2020).
\bibitem{qah3} J. Ge, Y. Liu, J. Li, H. Li, T. Luo, Y. Wu, Y. Xu and J. Wang, Natl
Sci. Rev. 7, 1280 (2020).
\bibitem{qi2} X.-L. Qi, T. L. Hughes and S.-C. Zhang, Phys. Rev. B 78,
195424 (2008).
\bibitem{fu2} L. Fu and C. L. Kane, Phys. Rev. B 76, 045302 (2007).
\bibitem{lcn1} A. M. Essin, J. E. Moore and D. Vanderbilt, Phys. Rev. Lett. 102, 146805
(2009).
\bibitem{lcn2} T. Rauch, T. Olsen, D. Vanderbilt and I.
Souza, Phys. Rev. B 98, 115108 (2018).
\bibitem{lcn3} N. Varnava1 and D. Vanderbilt, Phys. Rev. B 98,
245117 (2018).
\bibitem{ti} P. Hosur, S. Ryu and A. Vishwanath, Phys. Rev. B 81,
045120 (2010).
\bibitem{xuyong} P. Yang and Y. Xu, Phys. Rev. B 99, 195431 (2019).
\bibitem{fukui} T. Fukui, Y. Hatsugai and H. Suzuki, J. Phys. Soc. Japan
74, 1674 (2005).
\bibitem{bissard} J. Bellissard, arXiv:cond-mat/9504030.
\bibitem{hinge1} W. A. Benalcazar, B. A. Bernevig and T. L. Hughes,
Science 357, 61 (2017).
\bibitem{hinge2} W. A. Benalcazar, B. A. Bernevig and T. L. Hughes, Phys. Rev. B 96, 245115
(2017).
\bibitem{hinge3} F. Schindler, A. M. Cook, M. G. Vergniory, Z. Wang, S. S. P. Parkin, B. A. Bernevig and T.
Neupert, Sci. Adv. 4, eaat0346 (2018).
\bibitem{hinge4} Z. Song, Z. Fang and C. Fang, Phys. Rev. Lett.
119, 246402 (2017).
\bibitem{jpsj} Y. Hatsugai, J. Phys. Soc. Jpn. 75, 123601 (2006).
\bibitem{budich} J. C. Budich and E. Ardonne, Phys. Rev. B 88, 075419
(2013).
\bibitem{yan} Z. Yan, Z. Wu and W. Huang, Phys. Rev. Lett.
124, 257001 (2020).

\end{thebibliography}
\end{document}